\documentclass{article}

\usepackage{arxiv}

\usepackage[sort]{natbib}
\usepackage{graphicx}
\usepackage{ifthen}
\usepackage{amsmath,amssymb,amstext} 
\usepackage[ruled,vlined]{algorithm2e}
\usepackage[pdftex,pagebackref=false]{hyperref} 
\usepackage{comment}
\usepackage{color}
\RequirePackage{hyperref}
\usepackage{booktabs,siunitx}
\usepackage{graphics}

\usepackage[utf8]{inputenc} 
\usepackage[T1]{fontenc}    
\usepackage{hyperref}       
\usepackage{url}            
\usepackage{booktabs}       
\usepackage{amsfonts}       
\usepackage{nicefrac}       
\usepackage{microtype}      
\usepackage{lipsum}
\usepackage{graphicx}
\graphicspath{ {./images/} }

\usepackage{pgfplots}
\pgfplotsset{compat=1.18}
\usepackage{url}
\usepackage{amsmath}
\usepackage{amssymb}

\usepackage{caption}

\usepackage{multirow}
\usepackage{booktabs}
\usepackage{float}

\newcommand{\bJ}{\mbox{\bf J}}

\newcommand{\bZ}{\mbox{\bf Z}}
\newcommand{\bX}{\mbox{\bf X}}
\newcommand{\bx}{\mbox{\bf x}}

\newcommand{\bQ}{\mbox{\bf Q}}

\newcommand{\bb}{\mbox{\bf b}}

\newcommand{\bV}{\mbox{\bf V}}

\newcommand{\bW}{\mbox{\bf W}}

\newcommand{\bK}{\mbox{\bf K}}

\newcommand{\bO}{\mbox{\bf O}}

\def\bb{{\bf b}}

\newcommand{\bbeta}{\boldsymbol{\beta}}

\newcommand{\bepsilon}{\boldsymbol{\epsilon}}

\graphicspath{{Fig/}}

\title{Dynamic Survival Prediction using Longitudinal Images based on Transformer}

\author{
 Bingfan Liu \\
  Department of Statistics and Acturial Science\\
  Simon Fraser University\\
  Burnaby, BC, Canada \\
  \texttt{bingfan\_liu@sfu.ca} \\
   \And
 Haolun Shi \\
  Department of Statistics and Acturial Science\\
  Simon Fraser University\\
  Burnaby, BC, Canada \\
  \texttt{haoluns@sfu.ca} \\
   \And
 Jiguo Cao \thanks{Corresponding author.} \\
  Department of Statistics and Acturial Science\\
  Simon Fraser University\\
  Burnaby, BC, Canada \\
  \texttt{jiguo\_cao@sfu.ca} \\
}

\begin{document}

\maketitle
\begin{abstract}
Survival analysis utilizing multiple longitudinal medical images plays a pivotal role in the early detection and prognosis of diseases by providing insight beyond single-image evaluations. However, current methodologies often inadequately utilize censored data, overlook correlations among longitudinal images measured over multiple time points, and lack interpretability. We introduce SurLonFormer, a novel Transformer-based neural network that integrates longitudinal medical imaging with structured data for survival prediction. Our architecture comprises three key components: a Vision Encoder for extracting spatial features, a Sequence Encoder for aggregating temporal information, and a Survival Encoder based on the Cox proportional hazards model. This framework effectively incorporates censored data, addresses scalability issues, and enhances interpretability through occlusion sensitivity analysis and dynamic survival prediction. Extensive simulations and a real-world application in Alzheimer's disease analysis demonstrate that SurLonFormer achieves superior predictive performance and successfully identifies disease-related imaging biomarkers.
\end{abstract}


\section{Introduction}
Alzheimer’s disease (AD) is a progressively degenerative neurological disorder that poses major challenges to global healthcare due to its increasing prevalence and irreversible effects on cognition and overall quality of life. Projections indicate that by 2050, more than 100 million individuals worldwide are affected, which emphasizes the critical need for early and reliable methods to predict the onset of AD (\cite{brookmeyer2007forecasting, prince2015world}). In clinical practice, patients typically attend multiple hospital visits, resulting in sequences of MRI scans and biomarker measurements. Initiatives such as the Alzheimer’s Disease Neuroimaging Initiative (ADNI) compile extensive longitudinal MRI data to investigate disease progression (\cite{ibrahim2010basic, wang2024joint}). However, data censoring due to common issues such as dropout, loss of contact, or follow-up limitations further complicates accurate survival prediction.

Traditional joint modeling techniques for survival analysis, such as the FPCA joint Cox Regression Model (FPCA-Cox) and the Longitudinal FPCA joint Cox Regression Model (LoFPCA-Cox), integrate longitudinal biomarker trajectories with time-to-event data (\cite{andrinopoulou2018improved, rizopoulos2012joint, hsieh2006joint, shi2024dynamic}). Although FPCA-based methods suit sparse data well (\cite{yao2005functional}), they do not fully capture the complex, high-dimensional structure of MRI images and therefore risk missing critical global patterns. The biased estimators presented by the abovementioned two-stage modeling methods lead to inaccurate predictions on the survival outcome \citep{tsiatis2004joint, parr2022joint}.

At the same time, deep learning gains significant attention in medical imaging (\cite{litjens2017survey}), particularly in the context of survival prediction (\cite{katzman2018deepsurv, ranganath2016deep, luck2017deep, biganzoli1998feed}). For example, CNN-LSTM-based models (\cite{shu2021deep}) excel in extracting local features; however, they sometimes overlook global signals in MRI scans (\cite{lee2019dynamic}). Although LSTM layers capture temporal dependencies, challenges with interpretability and good prediction performance persist (\cite{giunchiglia2018rnn, ren2019deep}).

Transformer architectures, which feature multi-head self-attention mechanisms, demonstrate promise in spatiotemporal data analysis by effectively modeling both local and global relationships (\cite{xu2020spatial, du2022eeg, song2021transformer}). Despite this potential, the application of Transformers within survival models, especially for repeated imaging data collected over extended follow-up periods, remains underexplored (\cite{lee2019dynamic, luck2017deep}).

Some recent survival analysis studies (\cite{ren2019deep, lee2018deephit, lee2019dynamic, holste2024harnessing}) treat survival prediction as a multi-class classification problem optimized with cross-entropy loss, dividing event times into complementary bins as discrete labels. This formulation imposes an artificial independence among time bins, neglects the natural correlation between adjacent survival intervals, and potentially compromises predictive accuracy. Furthermore, a significant portion of real-world data is censored due to patient dropout, missed appointments, or limited follow-up durations. Relying exclusively on complete cases with cross-entropy loss discards valuable information and introduces bias.

In this work, we propose SurLonFormer, a Transformer-based joint modeling framework that simultaneously addresses the challenges of longitudinal imaging data and survival prediction. Our model achieves: \begin{itemize} \item Utilizing advanced multi-head attention mechanisms to capture global dependencies in longitudinal medical imaging data. \item Integrating multiple modalities, including scalar covariates and sequential images, to enhance both predictive performance and model flexibility. \item Employing a regularized negative log partial likelihood objective that leverages censored survival outcomes. \item Avoiding a multi-class classification formulation that ignores correlations among adjacent survival times. \item Offering interpretability through personalized dynamic survival probability estimates and occlusion sensitivity analysis that highlight patient-specific trajectories and disease-relevant imaging biomarkers. \end{itemize}

The remainder of this paper is organized as follows: Section Methodology details the proposed model; Section Simulation Studies describes the simulation study designed to evaluate model performance; Section Real Data Analysis presents the application of SurLonFormer to real-world data; and finally, Section Conclusion concludes the paper with a discussion of future research directions.

\section{Methodology}\label{section:methodology}
In this section, we present our methodology, beginning with a brief review of key survival analysis concepts. Consider a study observing a cohort of $I$ patients. Each patient $i$, for $i=0, \dots, I-1$, is associated with a set of covariates $\bx_i$.  For patient $i$, either the uncensored event occurrence time $U_i$, or the time of censoring, $C_i$ is recorded. The observed time at risk is defined as $ T_i = \min(U_i, C_i) $. A binary variable $\delta_i = I(U_i \le C_i)$ is used to indicate the occurrence of the event for patient $i$ where $ \delta_i = 1 $ signifies the occurrence of the event of interest and $\delta_i = 0 $ indicates censoring.

The objective of survival analysis is to estimate the distribution of event occurrence times, specifically
\begin{align*}
    \text{P}( U_i \le t) = \int^t_0 f(u) \text{d}u,
\end{align*}
where $f(u)$ is the corresponding probability density function. Instead of directly estimating this distribution, it is common practice to estimate the survival function, $S(t)$, defined as
\begin{align*}
    S(t) = \text{P}(U_i > t) = 1- \int^t_0 f(u) \text{d}u.
\end{align*}
To estimate $S(t)$, we model it through the hazard function $h(t)$, which is given by
\begin{align*}
    h(t) = \frac{f(t)}{S(t)} = \underset{\Delta t \rightarrow 0}{\text{lim}} \frac{1}{\Delta t} \text{P}( t \le U_i < t+\Delta t | U_i \ge t).
\end{align*}
The survival probability can then be expressed as
\begin{align*}
    S(t) = \text{exp}\Big\{ - H(t) \Big\} = \text{exp}\Bigg\{ - \int^t_0 h(u) du \Bigg\},
\end{align*}
where $H(t) = \int^t_0 h(u)du$ is the cumulative hazard function.

Under the CoxPH framework for survival analysis, the hazard function of patient $i$ is modeled as
\begin{align*}
    h(t) = \int^t_0 h_0(u)\text{exp}\{r_i(\bbeta)\}du,
\end{align*}
where $h_0(u)$ is the baseline hazard function. The risk score function $r_i(\bbeta) = f(\bbeta | \bx_i)$ of patient $i$ is usually estimated by a linear function $f(\cdot)$ with the parameter $\bbeta$ being the linear coefficient and  $\bx_i$ the covariate vector of patient $i$. Given the estimated model parameters, $\widehat \bbeta$, the baseline hazard function $h_0(t)$ can be estimated non-parametrically at time $t$ as
\begin{align*}
    \widehat h_0(t) = \frac{d_{(k)}}{\sum_{j \in R\{t_{(k)}\}} \text{exp}\Big\{r_j\big(\widehat \bbeta\big)\Big\}   },
\end{align*}
where $d_{(k)}$ is the number of events occurring at the distinct event occurrence time, $t_{(k)}$, ordered by the cumulated number of occurred events $k = 1, \cdots, \sum_i \delta_i$. $R\{t_{(k)}\}$ is the set of patients at risk for any time $t$ that $t_{(k-1)} \le t < t_{(k)}$ with $t_{(0)}$ being the study registration time. The cumulative baseline function $H_0(t) = \int^t_0 h_0(u)du$ can be approximated using Breslow's estimator as
\begin{align*}
    \widehat H_0(t) = \sum_{t_{(k)} \le t} \widehat h_0(t_{(k)}).
\end{align*}
 Thus, the probability of patient $i$ survived event-free after time $t$ is derived as
\begin{align}\label{survival_prob_eq}
    \widehat S(t) &= \text{exp}\Big[-\int^t_0 \widehat h_0(u)\text{exp}\{r_i (\widehat \bbeta)\}du\Big] \nonumber \\
           &= \text{exp}\Big[-\widehat H_0(t)\text{exp}\{r_i(\widehat \bbeta)\}\Big].
\end{align}

Estimation of model parameters, $\bbeta$, is performed by maximizing the likelihood function
\begin{align*}
L(\bbeta) &= \prod_{i=1}^{I}\Big[h_0(T_i)\text{exp}\{r_i(\bbeta)\}\Big]^{\delta_i}\text{exp}\Big[-H_0(T_i) \text{exp}\{r_i(\bbeta)\}\Big] \nonumber \\
                     &= L_1(\bbeta | \mathcal{D}) L_2\big\{\bbeta, h_0(\cdot) | \mathcal{D}\big\},\nonumber
\end{align*}
where $\mathcal{D}$ denotes the dataset and $L_1(\cdot)$ is the Cox partial likelihood function. Moreover, $L_2(\cdot)$ is the likelihood component mainly related to the baseline hazard and typically small which does not significantly influence the optimization process. By taking the log transformation, the objective function to maximize becomes
\begin{align}
\text{log} L_1(\bbeta) = \sum_{i=1}^{I} \delta_i \Bigg[r_i(\bbeta) - \text{log}\sum_{k \in R_i} \text{exp}\big\{r_i(\bbeta)\big\} \Bigg]. \nonumber
\end{align}

\subsection{Model Architecture}
We extend the traditional CoxPH framework to accommodate non-scalar and longitudinal data, specifically longitudinally collected medical images. Unlike the conventional CoxPH model which typically employs a linear predictor $f(\cdot)$, our approach leverages a more flexible architecture using a neural network to capture complex patterns in the data.

Consider a study where each patient undergoes multiple hospital visits till the end of the study at time $\tau$.  Let $J_i$ denote the maximum total number of visits for patient $i$, $i = 0, \cdots, I-1$. At each visit $j$, $0 \le j \le J_i-1$, an image of size $P \times P $ pixels is captured at time $t_{ij}$, represented as $\bX(t_{ij})$. Typically, medical images are grayscale, having a single color channel. To simplify notation, we treat each image as a two-dimensional matrix rather than a higher-dimensional tensor, implicitly embedding temporal information within the model. We further denote the event-free observation/landmark time as $t^*$ and the number of observed images till landmark time per event-free patient $i \in \{i | T_i \ge t^*, \delta_i = 0\}$ as $J^*$ where $J^* \le J_i$.

We propose SurLonFormer, a Transformer-based joint network designed for dynamic survival prediction using longitudinal medical images. The primary objective of SurLonFormer is to estimate the risk score $r_i$ for each patient $i$ and subsequently infer the distribution of event occurrence times through the survival function $S(t)$, as defined in Equation \ref{survival_prob_eq}. Our model comprises three main components: a vision encoder, a sequence encoder and a survival encoder. 

\subsubsection{Transformer Encoder Layer}
\begin{figure}
    \centering
    \includegraphics[width=0.85\textwidth]{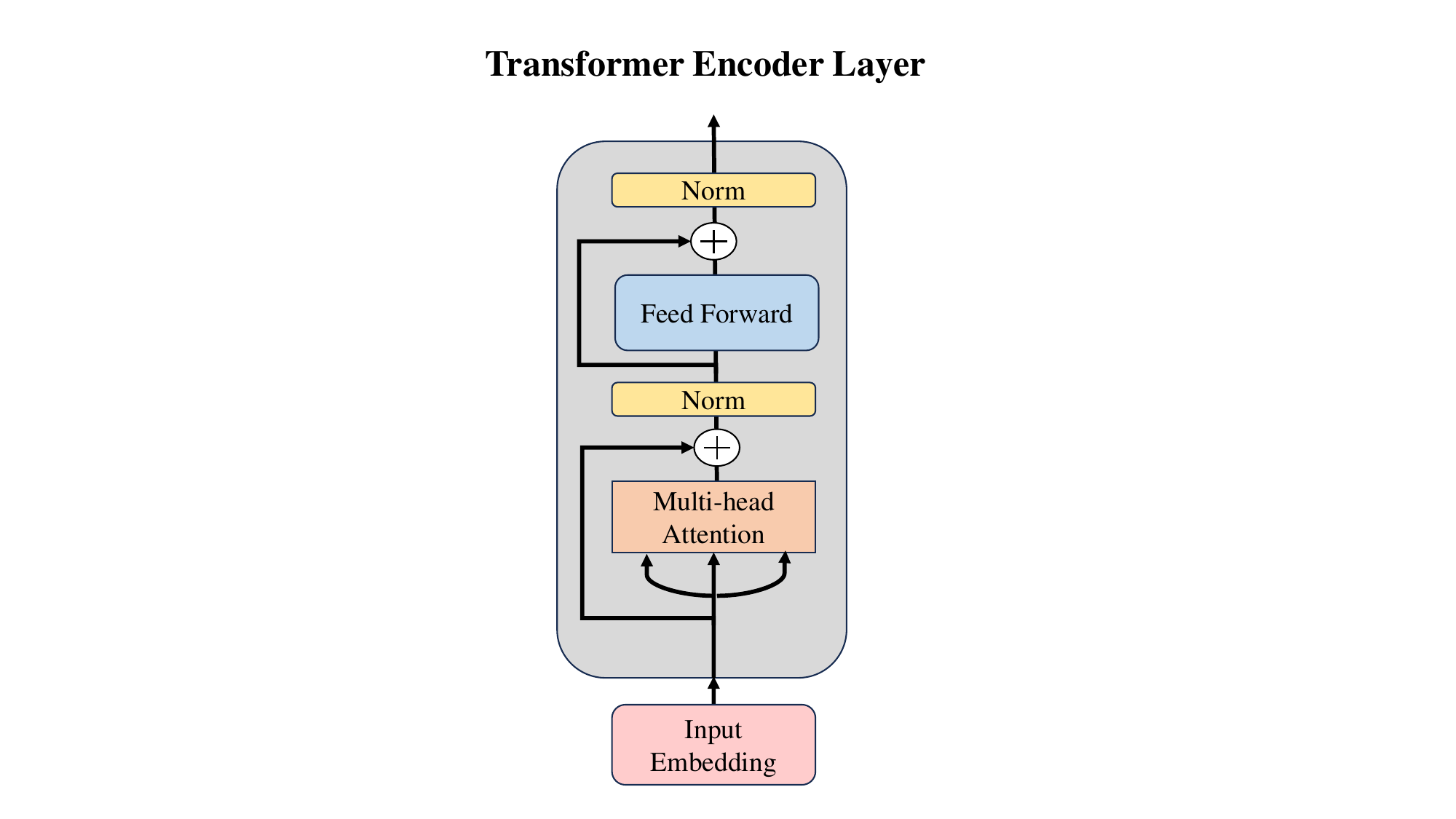}
    \caption{The Architecture of a Transformer Encoder block.}
    \label{transformer_encoder_layer}
\end{figure}

The fundamental building block of our model is the Transformer encoder (\cite{vaswani2017attention}) which consists of multi-head attention, a feed-forward neural network (FFN), residual connections, and layer normalization, as illustrated in Figure \ref{transformer_encoder_layer}.
The self-attention mechanism enables the model to capture relationships between its different inputs by computing the relevance of each patch relative to others. The multi-head variant enhances this capability by allowing the model to focus on various types of relationships simultaneously through multiple parallel attention heads.

Consider an input vector $\bZ$ of size $d$-by-1 for illustration. Given a predefined number of heads, $H$, Query, $\bQ_h$, Key, $\bK_h$, and Value, $\bV_h$, matrices are computed for each head $h$ for $h = 0, \dots, H-1$ as 
\begin{align*}
\bQ_h &= \bZ^\top\bW_{Qh}, \nonumber \\
\bK_h &= \bZ^\top\bW_{Kh}, \nonumber \\
\bV_h &= \bZ^\top\bW_{Vh}, \nonumber
\end{align*}
where $\bW_{Qh}, \bW_{Kh}, \bW_{Vh} \in \mathbb{R}^{d \times d_h}$ are the learnable weight matrices for head $h$ and $d_h = d/H$. 

The attention output for the head $h$, denoted as $\text{head}_h$, is computed using the scaled dot-product of query and key
\begin{align*}
\text{head}_h = \text{softmax}\Bigg(\frac{\bQ_h \bK_h^\top}{\sqrt{d_h}}\Bigg)\bV_h.\nonumber
\end{align*}
Multiple attention heads are computed in parallel. To simplify notation, let $\bQ, \bK$ and $\bV$ represent the concatenated Query, Key, and Value matrices across all heads, respectively. The outputs of all heads are then concatenated and linearly transformed to produce the multi-head attention output $\text{MultiHead}$ as
\begin{align*}
    \text{MultiHead} &= \text{Concat}\left\{ \text{head}_1, \dots, \text{head}_H \right\} \mathbf{W}_A,
\end{align*}
where $\bW_A \in \mathbb{R}^{d \times d}$ is a learnable parameter matrix.

The multi-head attention output is added to the original input $\bZ$ through a residual connection and then normalized as
\begin{align*}
\bO_r = \text{LayerNorm}\Bigg[\bZ + \text{MultiHead}\Big\{\bQ, \bK, \bV\Big\}\Bigg]. \nonumber
\end{align*}
A feed-forward neural network (FFN) then processes $\bO_r$ as follows:
\begin{align*}
\text{FFN}(\bO_r) = \text{GELU}\Big[\bO_r\bW_1 + \bb_1\Big]\bW_2 + \bb_2, \nonumber
\end{align*}
where GELU is the Gaussian Error Linear Unit activation function and $\bW_1 \in \mathbb{R}^{d \times d_{ff}}$ and $\bW_2 \in \mathbb{R}^{d_{ff} \times d}$ are learnable weight matrices with $d_{ff}$ be the hidden dimension of the feed-forward layer. The bias vectors $\bb_1$ and $\bb_2$ are in $\mathbb{R}^{P \times d_{ff}}$ and $\mathbb{R}^{d \times d_{ff}}$, respectively.

The output of the FFN is added to $\bO_r$ through another residual connection and normalized to produce the final output $\bO$ of the Transformer encoder layer
\begin{align*}
\bO = \text{LayerNorm}\{\bO_r + \text{FFN}(\bO_r)\}. \nonumber
\end{align*}

\subsubsection{Vision Encoder}
\begin{figure}
    \centering
    \includegraphics[width=0.85\textwidth]{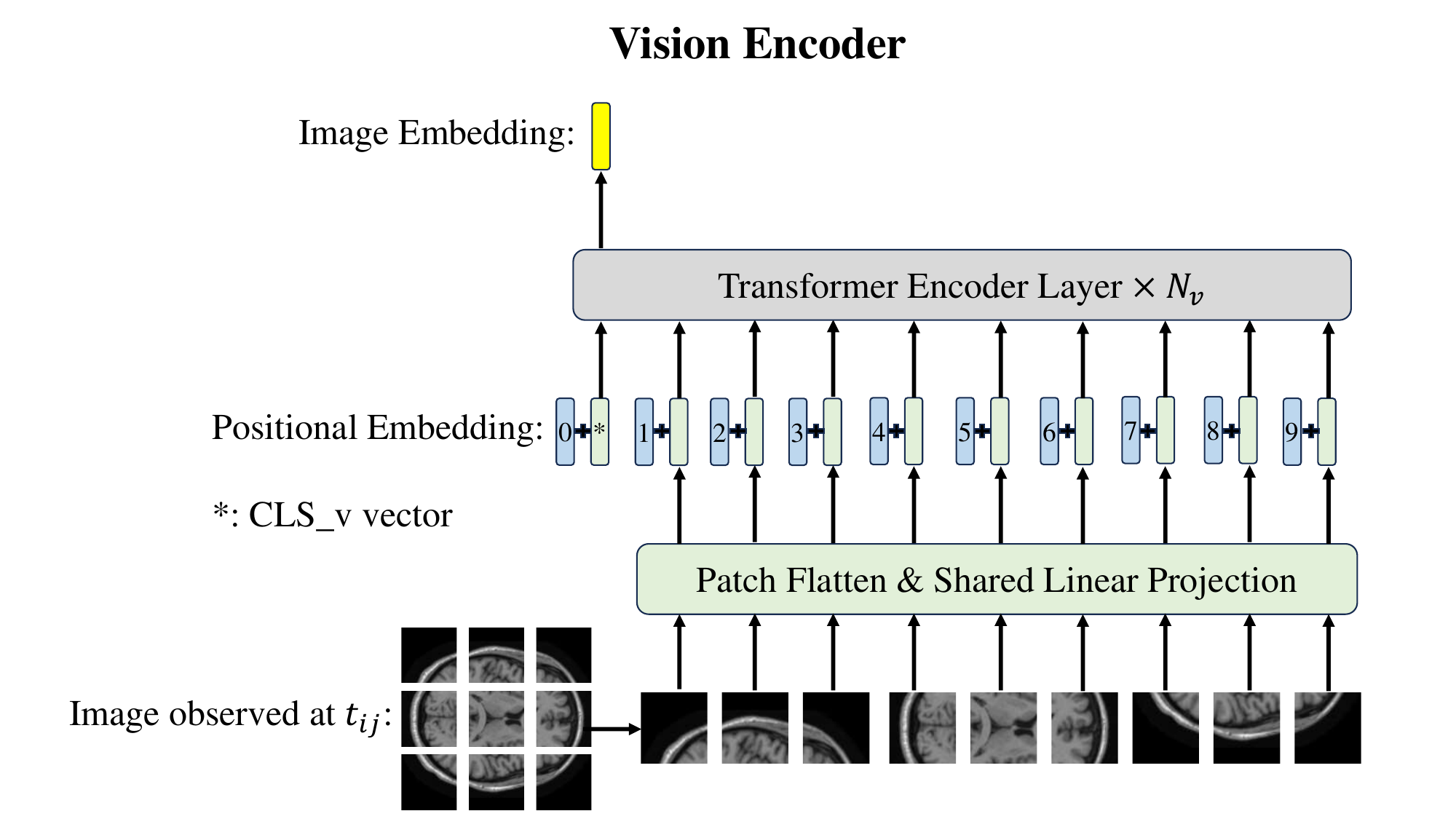}
    \caption{Model architecture for Vision Encoder of SurLonFormer.}
    \label{vision_encoder}
\end{figure}

The Vision Encoder is a Transformer-based network that encodes each of the $J^*$  images of the event-free patient $i$ into their respective embeddings, as shown in Figure \ref{vision_encoder}.

Borrowing the idea introduced by (\cite{alexey2020image}), each image is divided into $P$ equal-sized $\sqrt{P}$-by-$\sqrt{P}$ square patches. If the shape of an observed image is not suitable for division, bilinear interpolation is performed to reshape it into a square that can be divided into $P$ equal-sized $\sqrt{P}$-by-$\sqrt{P}$ patches. These patches are further flattened into vectors of size $P$-by-1 and linearly projected. The linear projection, $D(\cdot)$, involves an inner product with a learnable parameter matrix to create new features by projecting the flattened patches into a new feature space of size $d$. However, due to the limited data nature of medical data analysis and the fact that not every pixel in the medical image scan is useful (e.g., black pixels in the background of an MRI scan), $D(\cdot)$ with a smaller projection dimension $d$ is chosen as a dimension reduction operation to reduce the dimensionality of the patches and prevent model overfitting. 

A learnable token $\textbf{CLS\_v}$ is concatenated with the $P$ projected image patches and added with a learnable position embedding matrix $\textbf{PE}$ of size $(P+1)$-by-$d$ to encode their positional relationships within the image. This results in $\bZ_\text{pos}(t_{ij}) \in \mathbb{R}^{(P+1) \times d}$, where
\begin{align*}
&\bZ_{\text{pos}}(t_{ij}) = \bZ(t_{ij}) + \textbf{PE},\nonumber\\
&\bZ(t_{ij}) = \Big[\textbf{CLS\_v}, D\{\bX(t_{ij0})\} \dots, D\{\bX(t_{ijP-1})\}\Big]^\top \in \mathbb{R}^{(P+1) \times d} 
 \nonumber.
\end{align*}
Furthermore, $\bZ_\text{pos}(t_{ij})$ is passed through $N_v \ge 1$ Transformer encoder layers, $f^{(n)}_{\text{vt}}(\cdot)$ for $n=1,\cdots,N_v-1$ as
\begin{align*}
\bO^{(n)}_v(t_{ij}) &= f^{(n)}_{\text{vt}}\Big\{\bO^{(n-1)}_v(t_{ij})\Big\},
\end{align*}
where $\bO^{(n)}_v(t_{ij})$ represents the output matrix from the $n$-th transformer encoder layer for $0 \le n \le N_v-1$ with
\begin{align*}
\bO^{(0)}_v(t_{ij}) &= f^{(0)}_{\text{vt}}\Big\{\bZ_{\text{pos}}(t_{ij})\Big\}.
\end{align*}
Define the Vision Encoder as $f_v(\cdot)$. The image embedding $\bO_v(t_{ij})$ of the $j$-th image of patient $i$ observed at time $t_{ij}$, $\bX(t_{ij})$, is computed as
\begin{align*}
\bO_v(t_{ij}) = f_v\{\bX(t_{ij})\} = \Big\{\bO^{(N_v-1)}_v(t_{ij})[0,:]\Big\}^\top \in \mathbb{R}^{d \times 1}, \nonumber
\end{align*}
where $\bO^{(N_v-1)}_v(t_{ij})[0,:]$ represents the first row of the output matrix from the last transformer encoder layer, correspondingly to $\text{CLS\_v}$. 

As observed from the computation of the Transformer encoder layer, this method extracts image features globally through the attention mechanism. When the signals related to the disease of interest are scattered across the image, the attention mechanism can attend to image features that are far apart, unlike the receptive field in a CNN, which focuses only on local features within the convolution kernel.

\subsubsection{Sequence Encoder}
All image embeddings, $O_v(t_{ij})$, from the Vision Encoder for $j=0,\cdots,J^*-1$, of patient $i$ are concatenated with a special learnable $d$-by-$1$ vector, $\textbf{CLS\_l}$, as the input to the Sequence Encoder
\begin{align*}
\bO_{v,i}(t) = [\bO_v(t_{i0}), \dots, \bO_v(t_{iJ^*-1}), \textbf{CLS\_l}]^\top \in \mathbb{R}^{(J^*+1) \times d}.
\end{align*}
Furthermore, $\bO_{v,i}(t)$ is passed through $N_l \ge 1$ Transformer encoder layers, $f^{(n)}_{\text{lt}}(\cdot)$ for $n=1,\cdots, N_l-1$, to encode the disease progression information among the images of patient $i$ as
\begin{align*}
\bO^{(n)}_{l,i}(t) &= f^{(n)}_{\text{lt}}\Big\{\bO^{(n-1)}_{l,i}(t)\Big\},
\end{align*}
where $\bO^{(n)}_{l,i}(t)$ represents the output matrix from the $n$-th Transformer encoder layer for $0 \le n \le N_l-1$ with
\begin{align*}
\bO^{(0)}_{l,i}(t) &= f^{(0)}_{\text{lt}}\Big\{O_{v,i}(t)\Big\}.
\end{align*}
Unlike the Vision Encoder, a causal mask is incorporated into the attention mechanism of the Sequence Encoder. This causal mask ensures that the attention computation for each image in the sequence can only consider images at previous time points and itself (as well as the special $\textbf{CLS\_l}$ token), thereby respecting the temporal order.

\begin{figure}
    \centering
    \includegraphics[width=0.85\textwidth]{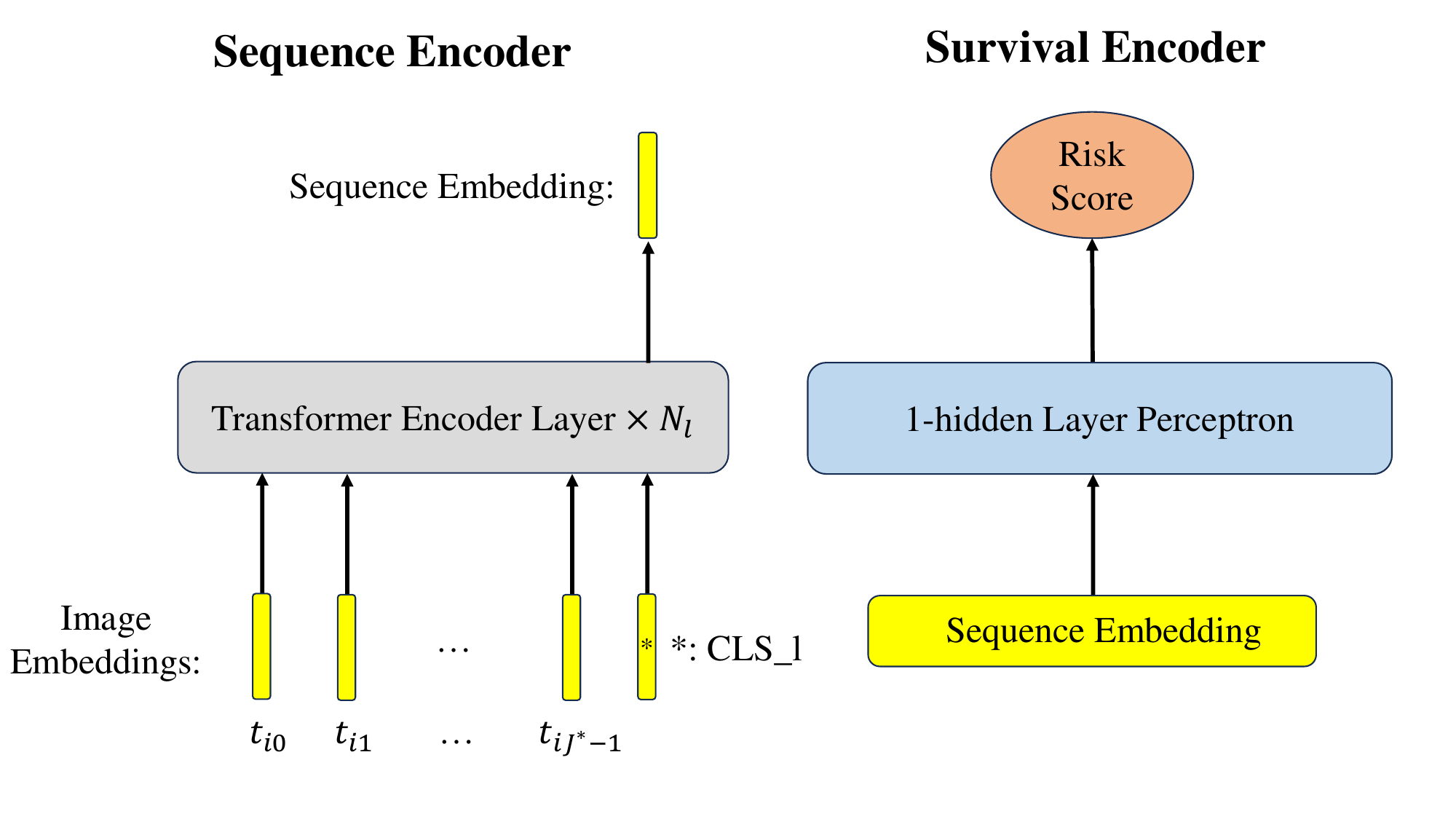}
    \caption{Model architecture for the Sequence Encoder and Survival Encoder of SurLonFormer.}
    \label{sequence_survival_encoder_combined}
\end{figure}

Define the Sequence Encoder as $f_l(\cdot)$. The sequence embedding, $\bO_{l,i}(t)$, of patient $i$ is computed as
\begin{align*}
\bO_{l,i}(t) = f_l\{\bO_{v,i}(t)\} = \Big\{\bO^{(N_l-1)}_{l,i}(t)[-1,:]\Big\}^\top \in \mathbb{R}^{d \times 1},\nonumber
\end{align*}
where $\bO^{(N_l-1)}_{l,i}[-1,:]$ is the last row of the output matrix $\bO^{(N_l-1)}_{l,i}(t)$, containing the summarized sequence information from the last Transformer encoder layer, corresponding to $\text{CLS\_l}$. The architecture of the Sequence Encoder is shown on the left side of Figure \ref{sequence_survival_encoder_combined}.

\subsubsection{Survival Encoder}
The Survival Encoder, $f_s(\cdot)$, shown on the right side of Figure \ref{sequence_survival_encoder_combined}, consists of a one-hidden-layer perceptron with GELU activation. It is utilized to estimate the risk score $r_i$ of patient $i$. In addition to receiving the sequence embedding $\bO_{l,i}(t)$ as input, the Survival Encoder can incorporate a scalar covariate vector $\bx_i \in \mathbb{R}^{d_x \times 1}$, such as the patient's age, sex, or other scalar-valued biomarkers. Denote the hidden layer dimension as $d_s$. The risk score is estimated as
\begin{align*}
r_i = f_s(\bO_{l,i}(t), \bx_i) = \text{GELU}\Big([\bO_{l,i}(t)^\top, \bx^{\top}_i]\bW_{s,1} + \bb_{s,1}\Big)\bW_{s,2} + b_{s,2},
\end{align*}
where $\bW_{s,1} \in \mathbb{R}^{(d+d_x) \times d_s}$ and $\bb_{s,1} \in \mathbb{R}^{1 \times d_s}$ are the weight matrix and bias vector in the hidden layer, and $\bW_{s,2} \in \mathbb{R}^{d_s \times 1}$ and $b_{s,2} \in \mathbb{R}$ are the weight vector and bias in the output layer. 

\begin{figure}
    \centering
    \includegraphics[width=1\textwidth]{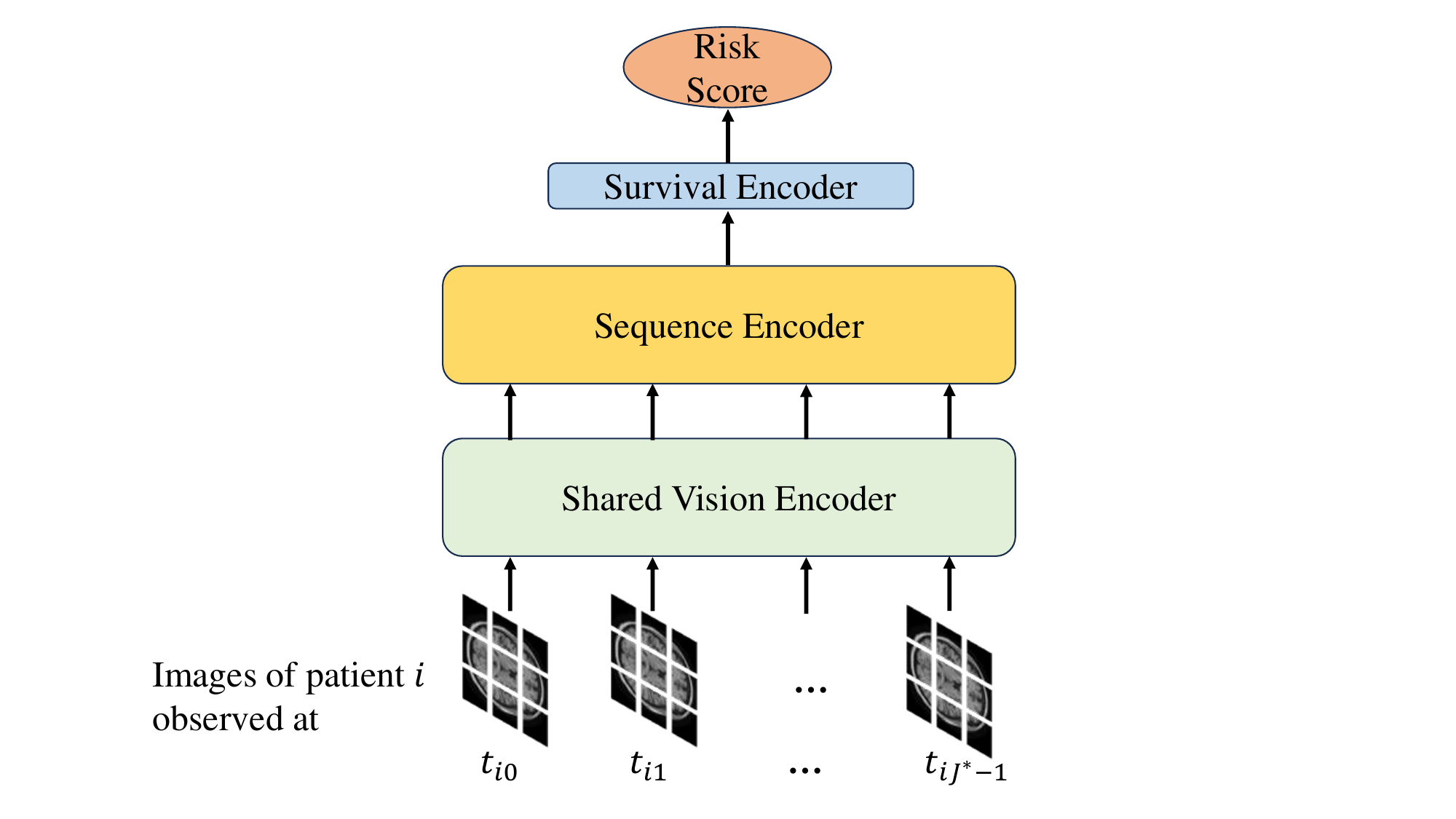}
    \caption{Model architecture for SurLonFormer.}
    \label{model_architecture_total}
\end{figure}

In total, the risk score of patient $i$ is computed as
\begin{align}\label{risk_score}
    r_i = f_s\Bigg\{f_l\Big[f_v\{\bX(t_{i,0})\}, \cdots, f_v\{\bX(t_{i,J^*-1})\}\Big], \bx_i \Bigg\}.
\end{align}
The survival probability can be computed conventionally as specified in Equation \ref{survival_prob_eq}. The overall architecture of the model is depicted in Figure \ref{model_architecture_total}.

\subsection{Computational Complexity and Memory Usage}
Each Transformer encoder layer has a self-attention module whose complexity scales as $\mathcal{O}(H \times n_\text{seq}^2 \times d)$, where $n_\text{seq}$ is the input sequence length. For the Vision Encoder, the sequence length is $(P+1)$, leading to a per-layer complexity of $\mathcal{O}\bigl(H \times (P+1)^2 \times d\bigr)$. Since there are $N_v$ layers in the Vision Encoder, the total complexity for the Vision Encoder is $\mathcal{O}\bigl(N_v \times H \times (P+1)^2 \times d\bigr)$.For the Sequence Encoder, the sequence length is $(J^*+1)$, leading to a per-layer complexity of $\mathcal{O}\bigl(H \times (J^*+1)^2 \times d\bigr)$. Hence, the total complexity of the Sequence Encoder is $\mathcal{O}\bigl(N_l \times H \times (J^*+1)^2 \times d\bigr)$. The Survival Encoder, being a small fully connected network, adds negligible overhead compared to the Vision and Sequence Encoders.

The memory footprint is dominated by the storage of embeddings and attention matrices in the Vision and Sequence Encoders. For the Vision Encoder, the token embeddings require storing $(P+1)\times d$ values per layer, and the self-attention module requires intermediate matrices of size $(P+1) \times (P+1)$. Similarly, for the Sequence Encoder, the token embeddings require $(J^*+1)\times d$ values, and the attention maps are of size $(J^*+1)\times (J^*+1)$. Thus, the overall memory usage is on the order of $\mathcal{O}\Big[N_v \big\{ d(P+1) + (P+1)^2\big\} + N_l \big\{ d(J^*+1)  + (J^*+1)^2\big\}\Big]$, with additional constants related to the number of attention heads $H$ and feed-forward dimensions. Overall, SurLonFormer is designed to balance flexibility and scalability by carefully choosing $N_v$, $N_l$, $d$, and $H$ based on the dataset size and available computational resources.

\subsection{Objective Function}
Adopting the CoxPH modeling framework offers several advantages: the log partial likelihood function is smooth, differentiable, and locally concave around its maximum, which facilitates optimization. Additionally, it allows for the estimation of model parameters without the need to explicitly estimate the baseline hazard function. This enables the model to focus solely on assessing the risk score, thereby simplifying the optimization process.

To estimate the model parameters, our method minimizes the regularized negative log partial likelihood using Elastic Net regularization. The regularizer is employed to prevent overfitting, which is commonly observed due to limited data. Thus, the objective function is defined as
\begin{align}
    l = \frac{1}{n_I}\sum_{i=1}^{I} \delta_i\Bigg\{\text{log}\sum_{k \in R_i} \text{exp}(r_k) -  r_i \Bigg\} + \lambda\Big\{\alpha||\bbeta ||_1 + (1-\alpha)||\bbeta ||_2 \Big\},\nonumber
\end{align}
where $n_I = \sum^I_i \delta_i$ and $r_i = f_s\Bigg\{f_l\Big[f_v\big\{\bX(t_{i,0})\big\}, \cdots, f_v\big\{\bX(t_{i, J^*-1})\big\}\Big]\Bigg\}$. By tuning regularization parameters $\lambda$ and $\alpha$, the trade-off between model complexity and fit can be balanced, thereby mitigating overfitting.

\subsection{Interpretability}
\subsubsection{Occlussion Sensitivity}
Occlusion sensitivity is employed to identify regions in the image that significantly influence the predicted risk score. To achieve this, we divide the entire image into multiple non-overlapping small regions. In each iteration, one small region in each image of the selected patient is masked with a baseline image patch, typically a black background. The corresponding change in the absolute value of the estimated risk score is recorded as the occlusion sensitivity of that region. By visualizing these sensitivity values in a 2D map and overlaying them on the original image, we can highlight the areas that are most important or indicative of disease, thereby providing insights into the model's decision-making process.

\subsubsection{Dynamic Survival Prediction}
For a new patient $i^*$ who is event-free up to time $t^* \in [0, \tau]$ with a set of longitudinally observed images $\bX(t_{i^*0}),\dots, \bX(t_{i^*J^*-1})$, where $t_{i^*J^*} \leq t^*$, the estimated risk score $\widehat{r}_{i^*}$ is computed using Equation \ref{risk_score}. Denote $U_{i^*}$ as the event occurrence time of subject $i^*$ and let $\Delta t > 0$ be a fixed time increment. The probability that subject $i^*$ survives beyond time $t^* + \Delta t < \tau$, given that the subject has survived event-free up to time $t^*$, is predicted as
\begin{align}\label{dynamic_survival_equation}
P(U_{i^*} > t^* + \Delta t | U_{i^*} > t^*) &= \frac{P(U_{i^*} > t^* + \Delta t )}{P(U_{i^*} > t^*)}\nonumber\\
    &= \frac{S(t^* + \Delta t)}{S(t^*)}\nonumber\\
    &= \frac{\exp\{-\int_0^{t^* + \Delta t}h(u)\text{d}u\}}{ \exp\{ -\int_0^{t^*}h(u)\text{d}u \}}\nonumber\\
    &= \Bigg\{\frac{S_{0}(t^* + \Delta t)}{S_{0}(t^*)}\Bigg\}^{\exp(\widehat{r}_{i^*})},
\end{align}
where $S_{0}(t)$ is the baseline survival function of subject $i^*$.

Dynamic survival prediction provides granular insights into how patients' survival probabilities evolve over time as the event-free landmark time increases. Moreover, in scenarios where the subset of event-free patients defined by $t^*$ is small, especially as $t^*$ approaches higher values, model training becomes challenging due to limited data. Dynamic survival prediction offers flexible survival forecasting based on Equation \ref{dynamic_survival_equation}, enabling the model to provide meaningful predictions even in data-scarce situations.

\subsection{Comparable Methods}
We implement several comparable methods based on existing models from the literature to evaluate the performance of SurLonFormer. Specifically, we implement two statistical methods, FPCA-Cox and LoFPCA-Cox 
and a modified deep learning method (\cite{lee2019dynamic}) that combines CNN and LSTM networks for comparison.

\subsubsection{FPCA-Cox}
FPCA is widely used in statistics for analyzing data recorded on a continuum, such as time or space. The FPCA-Cox method averages the image sequence of each patient and transforms each averaged image into a one-dimensional signal. FPCA is then applied to extract information from these signals.

Let $p \in \Theta_p = [0, N_p)$ denote the index of an image pixel, where $N_p$ is the total number of pixels in an image. The averaged image signal of patient $i$, $X_{i,\text{avg}}(p)$, is assumed to be a square-integrable stochastic process with a mean function $\mu(p)$ and a covariance operator $C(p, p')$:
\begin{align*}
    \mu(p) &= \text{E}\Big\{X_{i,\text{avg}}(p)\Big\},\\\nonumber
    C(p,p') 
           &= \text{E}\Bigg[\Big\{X_{i,\text{avg}}(p) - \mu(p)\Big\} \Big\{X_{i,\text{avg}}(p') - \mu(p')\Big\}\Bigg]\nonumber.
\end{align*}
Through the Karhunen-Loève expansion, the signal can be decomposed as
\begin{align*}
    X_{i,\text{avg}}(p) &= \mu(p) + \sum^{\infty}_{k=0} \xi_{ik} \phi_k(p)\\\nonumber
    &\approx \mu(p) + \sum^{N_k}_{k=0} \xi_{ik}\phi(p)\nonumber,
\end{align*}
where $N_k$ is the selected number of functional principal components (FPCs) for approximation. The FPCs, $\phi_k(p)$, form an orthonormal basis of the covariance operator, and the principal component scores are given by $\xi_{ik} = \int_{\Theta_p} \{X_{i,\text{avg}}(p) - \mu(p)\}\phi_k(p) \text{d}p$ for $k = 0, \cdots, N_k-1$. The extracted PC scores contain meaningful information along the dimensions of their corresponding FPCs and are used as covariates in the CoxPH model for survival analysis. Event-free survival prediction is then performed using Equation \ref{survival_prob_eq}.

\subsubsection{LoFPCA-Cox}
LoFPCA-Cox elevates the FPCA-Cox by modeling the entire image sequence of each patient using FPCA combined with functional linear mixed-effect method. This method extracts information from both the images and their progression over time. For patient $i$ observed at time $t_{ij}$, the image $X_{ij}(p)$ can be expanded as
\begin{align*}
    X_{ij}(p) &= \mu(p, t_{ij}) + X^0_{i}(p) + X^1_{i}(p)t_{ij} + W_{ij}(p)\nonumber,
\end{align*}
where $\mu(v, t_{ij})$ is the fixed main effect, $X_{i0}(p)$ is the functional random subject-specific intercept and $X_{i1}(p)$ is the functional random subject-specific time slope and $W_{ij}(p)$ is the functional random subject-visit-specific deviation. 

Assuming that $X_i(p) = [X^0_{i}(p), X^1_{i}(p)]^\top$ and $W_{ij}(p)$ are uncorrelated with their respective covariance operators $K^x(p, p')$ and $K^w(p, p')$, the signal can be further decomposed using the Karhunen-Loève expansion as
\begin{align*}
X_{ij}(p) &\approx \mu(p, t_{ij}) + \sum^{N_x}_{k_x=0} \xi_{ik_x}T^\top_{ij}\phi^x_{k_x}(p) + \sum^{N_w}_{x_w=0} \zeta_{ijk_w}\phi^w_{k_w}(p), \\\nonumber
\xi_{ik_x} &\stackrel{unc}{\sim} (0; \lambda^x_{k_x}), \\\nonumber
\zeta_{ijk_w} &\stackrel{unc}{\sim} (0; \lambda^w_{k_w}),\nonumber
\end{align*}
where $\phi^x_{k_x}(p) = [\phi^{x0}_{k_x}(p), \phi^{x1}_{k_x}(p)]^\top$ and $\phi^w_{k_w}(p)$ are the corresponding FPCs for the functional random subject-specific effect and subject-visit-specific deviation. $T_{ij} = [1,t_{ij}]^\top$. Their selected numbers of PC scores are respectively $N_x$ and $N_w$ with their corresponding PC scores to be $\xi_{ik}$ and $\zeta_{ijl}$. PC scores are uncorrelated and distributed (denoted as $\stackrel{unc}{\sim}$) with mean zero and variance $\lambda^x_{k_x}$ and $\lambda^w_{k_w}$ respectively. 

Similar to the FPCA-Cox method, the selected PC scores are used as covariates in the CoxPH model to perform dynamic survival prediction using Equation \ref{survival_prob_eq}. To ensure identifiability in parameter estimation using the functional linear mixed-effect model, some patients must have at least three images.

\subsubsection{CNN-LSTM}
A CNN-LSTM solution is implemented based on the approach proposed by  (\cite{lee2019dynamic}) for comparison. In this method, a CNN serves as the Vision Encoder, and an LSTM network acts as the Sequence Decoder. The CNN consists of several stacked blocks, each containing a 2D convolutional layer, batch normalization, Rectified Linear Unit (ReLU) activation, and dropout. The image embeddings produced by the CNN are fed into an LSTM network. Multiple LSTM layers can be stacked to capture temporal dependencies across the image sequence. The final hidden state from the LSTM serves as the sequence embedding. Similar to the proposed SurLonFormer, the Survival Encoder is a one-hidden-layer perceptron with GELU activation used to estimate the risk score of the patient. The model is optimized using the same objective function as in our proposed method. The number of layers, kernel sizes, and dropout rates are hyperparameters tuned during training. Dynamic survival prediction can be performed using Equation \ref{survival_prob_eq}, as described earlier.

\section{Simulation Studies}\label{section:simulation_study}
To assess the performance of SurLonFormer, we conduct a simulation study designed to evaluate the model's ability to accurately estimate patient risk and identify key regions in the images that influence risk prediction. In this study, we simulate data for $I = 700$ patients. The standardized ground truth image for patient $i$ at the registration time of the study, denoted as $\bX^{\text{gt}}(t_{i0})$, is a $64 \times 64$ 2D image with pixel values sampled from a uniform distribution, $\text{Unif}(-0.5, 0.5)$. Subsequently, one image is observed at each follow-up visit every six months, up to a maximum of 120 months (10 years) or until the occurrence of the event or censoring. For numerical stability, the time of the $j$-th visit for patient $i$, $t_{ij}$, is standardized to the unit interval, i.e., $t_{ij} \in [0,1]$ for all $i$ and $j = \{0, 1, \dots, J_i - 1\}$.

The sequence of ground truth images for patient $i$ observed at time $t_{ij}$ is defined as
\begin{align*}
\bX^{\text{gt}}_{ij} = \bX^{\text{gt}}_{i0} + 0.05t_{ij}\bJ_{64\times64} + 0.05t^2_{ij}\bJ_{64\times64}, \nonumber
\end{align*}
where $\bJ_{64\times64}$ is a 64-by-64 matrix of ones. The coefficient matrix $\bbeta$ is a 64-by-64 matrix with all entries set to zero except for eight 8-by-8 diagonal blocks. Each of these blocks is filled with values $g/70$ for $g = 0, \dots, 7$, representing the first block on the top-left corner to the last block on the bottom-right corner. The true risk score for subject $i$ is defined as
\begin{align}\label{risk_data_generation}
r_i = \sum_{j=0,1,2,3} \lVert\bbeta,\bX^{\text{gt}}_{ij}\rVert_F,
\end{align}
where $\lVert\cdot,\cdot\rVert_F$ denotes the Frobenius inner product. The hazard function for patient $i$ is defined as
\begin{align*}
    h_i(t) = h_0(t){\rm exp}\{ r_i \},
\end{align*}
with a constant baseline hazard function $h_0(t) = \exp(-5)$. 

To simulate the uncensored event occurrence time $U_i$ for patient $i$, we employ the inverse sampling method. We sample $u_i$ from a standard uniform distribution, $u_i \sim \text{Unif}(0,1)$ and compute $U_i$ as the inverse of the survival function at $u_i$, i.e., $U_i = S^{-1}(u_i)$, where $S(t)$ is the survival function defined in Equation \ref{survival_prob_eq}. The observed event occurrence time is $T_i = \min(U_i, 1)$, and the event indicator is $\delta_i = \mathbb{I}(T_i = U_i)$. Additionally, we introduce censoring by randomly censoring $5\%$ of the patients at any time during the study. Measurement error is incorporated by adding noise to the ground truth images. Let $\boldsymbol{\epsilon}(t_{ij})$ be a $64 \times 64$ matrix of noise sampled from a normal distribution, $\bepsilon(t_{ij}) \sim \mathcal{N}(0, 0.001)$. The observed image for patient $i$ at time $t_{ij}$ is then defined as
\begin{align*}
\bX(t_{ij}) = \bX^{\text{gt}}(t_{ij}) + \bepsilon(t_{ij})\nonumber.
\end{align*}
The simulation settings are designed such that the images are highly non-smooth with image features distributed globally within the main diagonal blocks. The images of a patient change non-linearly over time, and the coefficient matrix $\boldsymbol{\beta}$ is block-wise constant with non-smooth edges, making risk estimation challenging.

We perform a 4-fold cross-validation to evaluate the models. The models are trained on the training set and evaluated on the validation set at different landmark times $t^* \in \{12, 18, 24\}$ months and time increments $\Delta t \in \{12, 24, 48\}$ months. For each scenario, we compute the averaged time-dependent Area Under the Receiver Operating Characteristic Curve (AUC) (\cite{hung2010estimation}), time-dependent Concordance Index (C-index) (\cite{antolini2005time}), and time-dependent Brier Scores (BS) (\cite{blanche2015quantifying}), along with the true time-dependent AUC ($\text{true AUC}$) to assess the quality of the risk score predictions. The simulation is repeated for 100 runs, and the averaged performance measures are presented in Table \ref{simulation_table}.
\begin{table}[!ht]
\centering
\small
\setlength{\tabcolsep}{6pt}
\begin{tabular}{lllcccc}
\toprule
$t^*$ & $\Delta t$ & Method & True AUC & AUC & C-index & BS \\
\midrule
12 & 12 & FPCA-Cox & 0.809 & 0.554 & 0.549 & 0.111 \\
 &  & LoFPCA-Cox & 0.809 & 0.590 & 0.588 & 0.078 \\
 &  & CNN-LSTM & 0.809 & 0.609 & 0.604 & 0.117 \\
 &  & SurLonFormer & 0.809 & 0.772 & 0.755 & 0.107 \\
\addlinespace
12 & 24 & FPCA-Cox & 0.833 & 0.547 & 0.542 & 0.169 \\
 &  & LoFPCA-Cox & 0.833 & 0.590 & 0.584 & 0.145 \\
 &  & CNN-LSTM & 0.833 & 0.600 & 0.592 & 0.180 \\
 &  & SurLonFormer & 0.833 & 0.774 & 0.750 & 0.159 \\
\addlinespace
12 & 48 & FPCA-Cox & 0.815 & 0.543 & 0.537 & 0.231 \\
 &  & LoFPCA-Cox & 0.815 & 0.583 & 0.572 & 0.220 \\
 &  & CNN-LSTM & 0.815 & 0.591 & 0.579 & 0.243 \\
 &  & SurLonFormer & 0.815 & 0.774 & 0.733 & 0.199 \\
\addlinespace
18 & 12 & FPCA-Cox & 0.803 & 0.564 & 0.561 & 0.105 \\
 &  & LoFPCA-Cox & 0.803 & 0.595 & 0.591 & 0.079 \\
 &  & CNN-LSTM & 0.803 & 0.613 & 0.607 & 0.115 \\
 &  & SurLonFormer & 0.803 & 0.727 & 0.714 & 0.108 \\
\addlinespace
18 & 24 & FPCA-Cox & 0.802 & 0.542 & 0.542 & 0.151 \\
 &  & LoFPCA-Cox & 0.802 & 0.591 & 0.585 & 0.130 \\
 &  & CNN-LSTM & 0.802 & 0.611 & 0.602 & 0.169 \\
 &  & SurLonFormer & 0.802 & 0.746 & 0.722 & 0.148 \\
\addlinespace
18 & 48 & FPCA-Cox & 0.796 & 0.529 & 0.531 & 0.227 \\
 &  & LoFPCA-Cox & 0.796 & 0.577 & 0.568 & 0.218 \\
 &  & CNN-LSTM & 0.796 & 0.594 & 0.582 & 0.259 \\
 &  & SurLonFormer & 0.796 & 0.729 & 0.700 & 0.225 \\
\bottomrule
\end{tabular}
\vspace{0.5cm}
\caption{\footnotesize{True time-dependent AUC (true AUC) and averaged estimated time-dependent AUC scores (AUC), concordance index (C-index) and Brier scores (BS) from 4-fold cross-validation over 100 simulation runs of FPCA-Cox, LoFPCA-Cox, CNN-LSTM, and SurLonFormer (proposed method) at various landmark times ($t^*$) and time increments ($\Delta t$). Higher AUC and C-index values indicate better predictive performance, while lower Brier Scores (BS) denote improved calibration.}}
\label{simulation_table}
\end{table}

Higher values in the time-dependent AUC and C-index and a lower value of the time-dependent Brier Score indicate better model performance. As shown in Table \ref{simulation_table}, SurLonFormer outperforms all other methods across all scenarios, achieving the highest AUC and C-index scores, indicating its superior discriminating power. The consistently low Brier Scores indicate that SurLonFormer provides well-calibrated risk predictions. The CNN-LSTM model ranks second in performance, showing substantial improvements over the FPCA-based methods. However, it exhibits higher Brier Scores, suggesting some degree of overfitting. This overfitting likely arises because the CNN-LSTM architecture tends to focus on local image features, which may not capture the global distribution of disease-related features effectively. Both FPCA-Cox and LoFPCA-Cox methods demonstrate lower AUC and C-index values compared to CNN-LSTM and SurLonFormer, reflecting their limited ability to capture complex spatial and temporal dependencies in the image data. However, these methods maintain relatively lower Brier Scores, indicating better calibration despite their reduced discrimination capabilities.

\begin{figure}
    \centering
    \includegraphics[width=0.9\textwidth]{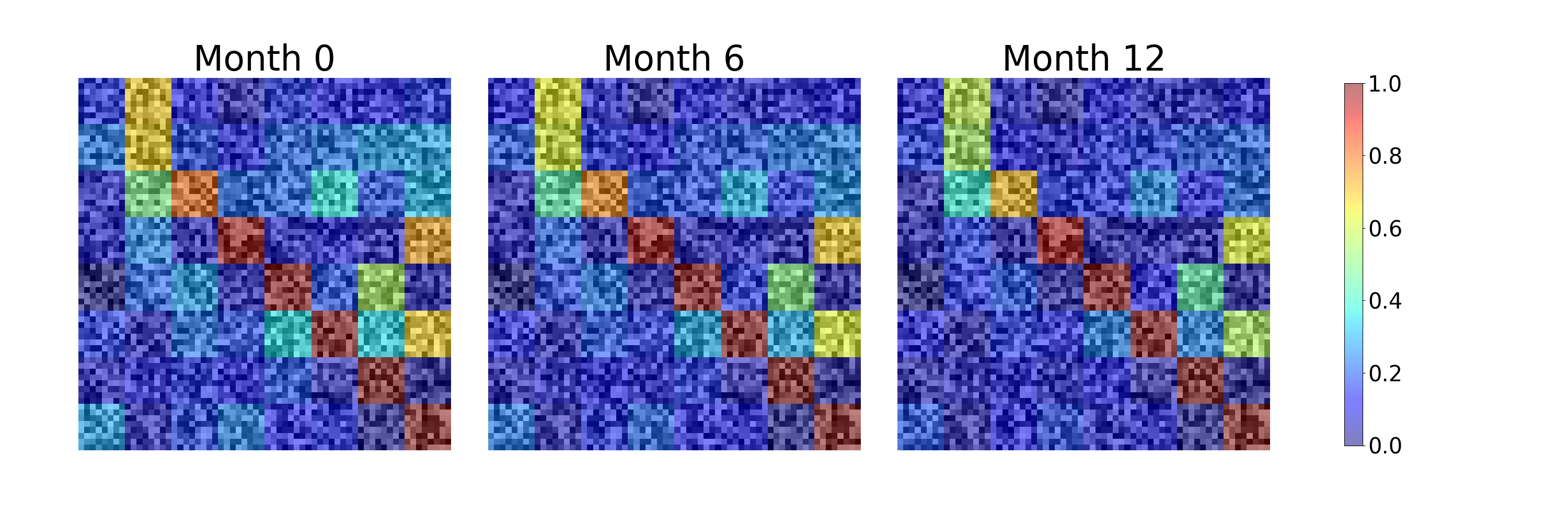}
    \caption{\footnotesize{Occlusion sensitivity maps overlaid on the observed images of a selected patient at Months 0, 6, and 12. Dark red regions indicate areas with higher sensitivity, signifying their greater influence on the risk score prediction. Blue regions indicate areas with lower sensitivity.}}
    \label{sim_os}
\end{figure}

To evaluate whether SurLonFormer can accurately identify regions in the image that influence the risk score, we perform occlusion sensitivity analysis on a selected patient. Specifically, we provide the model with the patient's first three observed images at Months 0, 6, and 12. The computed occlusion sensitivity maps are overlaid on the observed images and visualized in Figure \ref{sim_os}.
As observed in Figure \ref{sim_os}, each of the three images exhibits higher occlusion sensitivity (marked in dark red) within the majority of eight main diagonal 8-by-8 blocks. This aligns with the data-generating process, where these blocks contain the disease-related features contributing to the risk score, as defined in Equation (\ref{risk_data_generation}). The model successfully identifies most of the critical regions, demonstrating its capability to focus on meaningful areas within the images. Additionally, some off-diagonal regions show moderate sensitivity values ranging from 0.6 to 0.7. These may arise due to the high dropout rates implemented in the model to prevent overfitting. As the timestamp increases, the sensitivity values in these misidentified areas decrease, indicating improved localization of the risky regions as more image information becomes available. Overall, the occlusion sensitivity analysis confirms that SurLonFormer effectively identifies the relevant regions in the images that influence the risk predictions, validating both the model's interpretability and its alignment with the underlying data structure.

\section{Real Data Analysis}\label{section:real_data_analysis}
To illustrate the performance of SurLonFormer in a real-world scenario, we apply the proposed model and the comparable methods to the Alzheimer's Disease Neuroimaging Initiative (ADNI) database.

The ADNI database comprises longitudinally collected MRI scans of study patients every six months during the initial two years, followed by scans every twelve months until the end of their follow-up period, typically spanning over ten years, with occasional missed follow-ups. One MRI image is observed for each patient per visit. Image preprocessing is conducted using image processing pipelines based on FreeSurfer version 5.3.0 (\cite{ma2019quantitative, popuri2020using}) to prepare the T1-weighted structural brain MRI data. The images represent affine-registered volumetric scans of the middle anatomical region containing key AD-related functional brain areas. All scans are standardized to the same stereotaxic space and scale across all patients. During each visit, patients receive one of three diagnoses: cognitive normal (CN), mild cognitive impairment (MCI), or Alzheimer’s disease (AD). We consider the time until a patient's initial diagnosis of AD as the primary outcome. After excluding patients with improperly registered images, we focus on a cohort of 1,143 patients who were free of AD at baseline, of which 267 eventually progressed to AD. The average time at risk for developing the disease is 37.80 months with a standard deviation of 23.69 months.

For patients who remain event-free up to a landmark time $t^*$, we perform a 4-fold cross-validation. The models are trained on the training set and evaluated on the validation set. We assess the model performance across six different scenarios defined by landmark times $t^* \in \{12, 24\}$ months and time increments $\Delta t \in \{12, 18, 24\}$ months. 
Similar to the simulation study, we compute the averaged time-dependent AUC, Concordance Index (C-index) and Brier Score (BS). The averaged performance measures are presented in Table \ref{real_table}.
\begin{table}[!ht]

\centering
\tiny
\setlength{\tabcolsep}{9pt}
\resizebox{0.7\columnwidth}{!}{
\begin{tabular}{cclccc}
\toprule
$t^*$ & $\Delta t$ & Method & AUC & C-index & BS \\ 
\midrule
12 & 12 & FPCA-Cox       & 0.652 & 0.661 & 0.063 \\
 &  & LoFPCA-Cox     & 0.594 & 0.603 & 0.064 \\
 &  & CNN-LSTM       & 0.697 & 0.690 & 0.064 \\
 &  & SurLonFormer  & 0.726 & 0.725 & 0.065 \\
\addlinespace
12 & 18 & FPCA-Cox       & 0.628 & 0.635 & 0.100 \\
 &  & LoFPCA-Cox     & 0.584 & 0.585 & 0.101 \\
 &  & CNN-LSTM       & 0.712 & 0.697 & 0.095 \\
 &  & SurLonFormer  & 0.742 & 0.724 & 0.095 \\
\addlinespace
12 & 24 & FPCA-Cox       & 0.638 & 0.636 & 0.121 \\
 &  & LoFPCA-Cox     & 0.588 & 0.587 & 0.121 \\
 &  & CNN-LSTM       & 0.725 & 0.701 & 0.113 \\
 &  & SurLonFormer  & 0.732 & 0.719 & 0.114 \\
\addlinespace
24 & 12 & FPCA-Cox       & 0.564 & 0.565 & 0.032 \\
 &  & LoFPCA-Cox     & 0.715 & 0.708 & 0.031 \\
 &  & CNN-LSTM       & 0.651 & 0.649 & 0.031 \\
 &  & SurLonFormer  & 0.827 & 0.823 & 0.030 \\
\addlinespace
24 & 18 & FPCA-Cox       & 0.673 & 0.653 & 0.057 \\
 &  & LoFPCA-Cox     & 0.683 & 0.680 & 0.057 \\
 &  & CNN-LSTM       & 0.696 & 0.686 & 0.053 \\
 &  & SurLonFormer  & 0.777 & 0.772 & 0.052 \\
\addlinespace
24 & 24 & FPCA-Cox       & 0.580 & 0.604 & 0.092 \\
 &  & LoFPCA-Cox     & 0.583 & 0.607 & 0.092 \\
 &  & CNN-LSTM       & 0.733 & 0.710 & 0.085 \\
 &  & SurLonFormer  & 0.748 & 0.740 & 0.084 \\
\bottomrule
\end{tabular}
}
\vspace{0.5cm}
\caption{\footnotesize{Averaged estimated time-dependent AUC scores (AUC), concordance index (C-index) and Brier scores (BS) from 4-fold cross-validation of FPCA-Cox, LoFPCA-Cox, CNN-LSTM, and SurLonFormer (proposed method) at various landmark times ($t^*$) and time increments ($\Delta t$). Higher AUC and C-index values indicate better predictive performance, while lower Brier Scores (BS) denote improved calibration.}}
\label{real_table}
\end{table}

SurLonFormer consistently outperforms the other methods in terms of the time-dependent AUC and C-index across all scenarios, indicating superior discriminative ability. Additionally, SurLonFormer achieves the lowest Brier Scores in almost all scenarios, demonstrating excellent calibration. As $t^*$ and the number of observed images increase, the performance of SurLonFormer is consistently improved evidenced by higher AUC and C-index values and lower Brier Scores. This indicates an outstanding learning ability for SurLonFormer, and its prediction performance can be improved with more input information. In contrast, traditional methods like FPCA-Cox and LoFPCA-Cox show lower AUC and C-index values and higher Brier Scores, highlighting their limited ability to capture complex spatial and temporal dependencies in complex image data. The CNN-LSTM model performs better than the FPCA-based methods but still lags behind SurLonFormer. Overall, the results indicate that SurLonFormer provides more accurate and reliable survival predictions, effectively leveraging both spatial and temporal information from longitudinal medical images.

\begin{figure}
    \centering
    \includegraphics[width=0.8\textwidth]{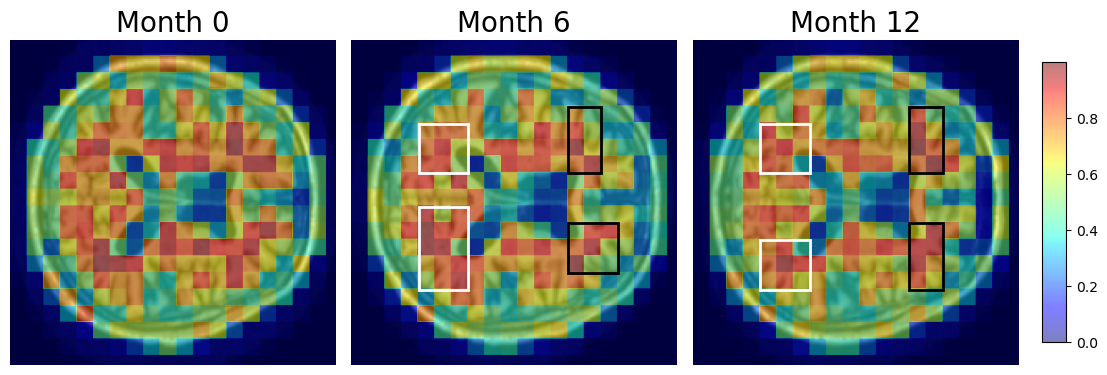}
    \caption{\footnotesize{Occlusion sensitivity maps overlaid on the observed images of patient $i^*$ at Months 0, 6, and 12. Darker red regions indicate higher sensitivity, signifying greater influence on the risk score prediction.}}
    \label{real_OS}
\end{figure}

To evaluate the interpretability of SurLonFormer, we perform occlusion sensitivity analysis on a selected patient. We choose a patient $i^*$ who remains event-free for 12 months since registration and develops AD by the end of the study. The patient's first three observed images at months 0, 6, and 12 are used for this analysis. The computed occlusion sensitivity maps are overlaid on the observed images and visualized in Figure \ref{real_OS}.
where the brighter and redder colors indicate higher sensitivity values, while darker and bluer colors represent lower occlusion sensitivity values. The right side of each subfigure corresponds to the anterior (front) of the brain. The left side of each subfigure corresponds to the posterior (back). Redder regions represent areas most strongly associated with Alzheimer’s disease pathology, while bluer regions indicate areas of lower relevance.

At Month 0, key regions of high sensitivity are observed broadly around the center of the brain. However, there is no significant visual indication of distinct red patch clusters, suggesting that the model has not yet specialized in identifying critical regions. By Month 6, certain blue regions expand and intensify, particularly in the black areas of the scan representing the background, indicating the model's increasing confidence in distinguishing non-critical regions. Concurrently, red regions begin to form distinct clusters, particularly within the frontal lobe (marked by black squares) and at the intersection of the temporal and occipital lobes (marked by white squares). These changes reflect the model's growing focus on disease-relevant areas. By Month 12, the figure reveals more pronounced clustering of sensitivity in these regions, further indicating the progression of Alzheimer’s disease-related changes. The background regions outside the brain consistently exhibit lower sensitivity values, as expected, reaffirming the model’s ability to disregard irrelevant features.

The identified cluster regions correspond to brain functionalities closely linked to Alzheimer’s disease. The frontal lobe is associated with intelligence, judgment, and behavior, while the temporal lobe is crucial for memory. The occipital lobe, responsible for vision, also shows involvement as the disease progresses. The temporal progression of sensitivity aligns with known patterns of Alzheimer’s disease pathology, where early involvement in the frontal and temporal lobes spreads to additional areas, such as the occipital lobe, as the disease advances. These results underscore the model’s ability to capture both the spatial and temporal dynamics of Alzheimer’s disease progression.

\begin{figure}
    \centering
    \includegraphics[width=\textwidth]{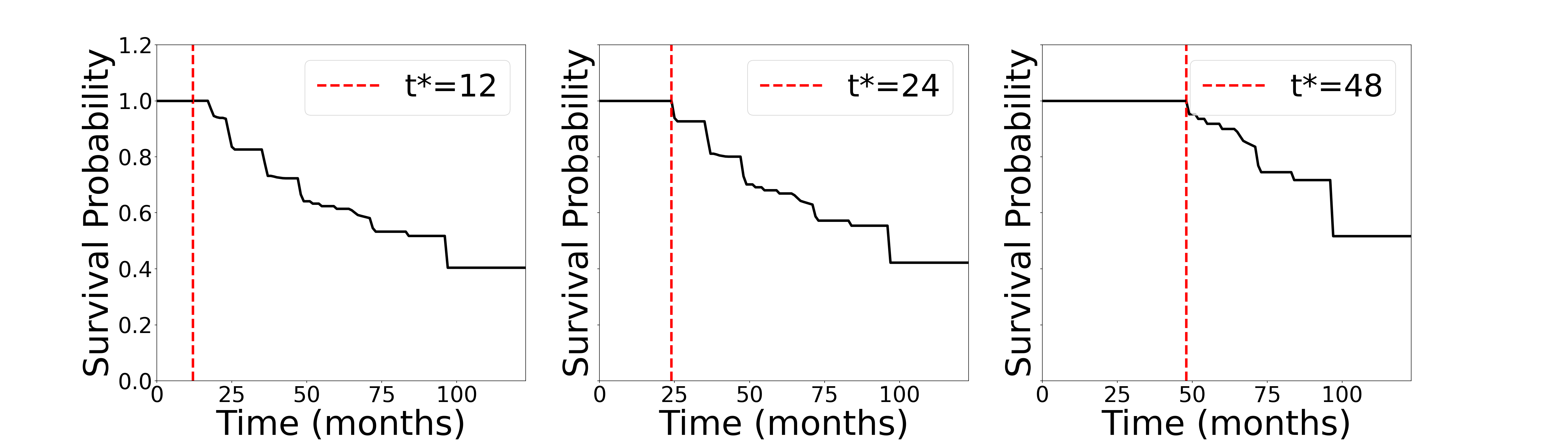}
    \caption{\footnotesize{Predicted dynamic survival probabilities of patient $i^*$ at landmark times $t^* = \{12, 24, 48\}$ months. The red dashed lines indicate the survival probabilities at each respective $t^*$.}}
    \label{real_survival_pred}
\end{figure}

To gain insights into the dynamic survival probabilities predicted by our model, we compute the survival probabilities for patient $i^*$ at different event-free landmark times $t^* = \{12, 24, 48\}$ months using Equation \ref{dynamic_survival_equation}. The predicted survival probabilities are visualized in Figure \ref{real_survival_pred}.
The dynamic survival probability plot provides valuable insights into individual patient trajectories and validates the clinical relevance of SurLonFormer’s predictions. In all scenarios, the survival probability of patient $i^*$ decreases as time progresses beyond the landmark time $t^*$, which is consistent with the eventual development of AD. At $t^*=12$, the survival probability begins high and gradually declines, reflecting the initial assessment of risk. As $t^*$ increases, the survival probability at shorter $\Delta t$ intervals remains high, suggesting a lower immediate risk of developing AD.  Patients who remain event-free for longer periods tend to exhibit higher survival probabilities, consistent with clinical expectations. However, it experiences a steep decline after. Particularly, when $t^* = 24$, the sharp decline happens after two years indicating the eventual disease occurrence. 

In summary, the real data analysis demonstrates that SurLonFormer not only outperforms traditional statistical methods like FPCA-Cox and LoFPCA-Cox but also surpasses the CNN-LSTM model in both predictive performance and interpretability. The high AUC and C-index values indicate that SurLonFormer effectively discriminates between patients who will develop AD and those who will not. Moreover, the low Brier Scores suggest that the model's predicted survival probabilities are well-calibrated. The occlusion sensitivity analysis further validates the model's interpretability by successfully identifying key brain regions associated with AD risk, consistent with established clinical findings. This capability is crucial for gaining trust in the model's predictions and for providing actionable insights in a clinical setting. Additionally, the dynamic survival probability predictions offer a personalized view of disease progression, enabling clinicians to monitor and adjust treatment plans based on individual risk trajectories.

\section{Conclusion}\label{section:conclusion}
In this paper, we propose a transformer-based joint model, SurLonFormer, designed to analyze longitudinally observed medical images and scalar covariates from structured datasets for dynamic survival prediction. The model leverages a vision encoder to extract image features, which are then aggregated over the image sequence using a sequence encoder. A survival encoder subsequently summarizes these features to estimate the risk of patients developing the disease. The Cox Proportional Hazards (CoxPH) framework is employed to optimize the model and predict patients' survival probabilities. Additionally, SurLonFormer inherently provides model interpretability through occlusion sensitivity analysis and dynamic survival prediction, enabling the identification of key regions in medical images that influence risk predictions and gaining insights into personalized disease progression trajectories.

Through extensive simulation experiments and a real-world application using the Alzheimer's Disease Neuroimaging Initiative (ADNI) database, SurLonFormer demonstrates superior performance compared to existing methods, including FPCA-Cox, LoFPCA-Cox, and CNN-LSTM models. The proposed model achieves higher time-dependent AUC and C-index values, along with lower Brier Scores, indicating better discrimination and calibration in survival predictions. On the other hand, when applied to rare diseases with even smaller datasets, the model's performance may decline due to overfitting and limited data availability. We leave these challenges for future research work.

\bibliographystyle{unsrt}

\end{document}